\newcommand{\beq}{\begin{equation}}
\newcommand{\eeq}{\end{equation}}
\newcommand{\vecr}{{\rm{\bf{r}}}}
\newcommand{\vecM}{{\rm{\bf{M}}}}

\def\etal{{\it et al.\/} }
\def\eg{{\it e.g.\/} }
\def\ie{{\it i.e.\/} }
\def\ROSAT{{\it ROSAT\/} }

\def\Einstein{{\it Einstein\/} }
\documentstyle[emulateapj]{article}
\submitted{accepted for publication in the Astrophysical Journal, 24 March 1999}

\lefthead{SULKANEN}
\righthead{GALAXY CLUSTER SHAPES AND SYSTEMATIC ERRORS IN H$_0$...}

\begin{document}

\title{Galaxy Cluster Shapes and Systematic Errors in H$_0$\\
Measured by the Sunyaev-Zel'dovich Effect}

\author{Martin E. Sulkanen\altaffilmark{1,2}}
\affil{X-Ray Astronomy Group, Space Sciences Laboratory, NASA/Marshall Space
Flight Center, Huntsville, AL 35812}

\altaffiltext{1}{Rackham Visiting Scholar, University of Michigan.} 
\altaffiltext{2}{present address: Department of Astronomy,
    University of Michigan, Ann Arbor, MI 48109-1090}

\begin{abstract}
Imaging of the Sunyaev-Zel'dovich (SZ) effect in galaxy clusters
combined with cluster plasma x-ray diagnostics can measure the
cosmic distance scale to high redshift. However, projecting the
inverse-Compton scattering and x-ray emission along the cluster
line-of-sight introduces systematic errors in the Hubble constant,
$H_0$, because the true shape of the cluster is not known.
In this paper I present a study of the systematic errors in the value
of $H_0$, as determined by the x-ray and SZ properties of theoretical
samples of triaxial isothermal ``beta'' model clusters, caused by projection
effects and observer orientation relative to the model clusters' principal
axes. I calculate three estimates for $H_0$ for each cluster, based on
their large and small apparent angular core radii, and their arithmetic
mean. I demonstrate that the estimates for $H_0$ for a sample of 25 clusters 
have limited systematic error: the 99.7\% confidence
intervals for the mean estimated $H_0$ analyzing the clusters using either 
their large or mean angular core radius are within
$\simeq 14\%$ of the ``true'' (assumed) value of $H_0$ (and enclose it), 
for a triaxial beta model cluster sample possessing a distribution of
apparent x-ray cluster ellipticities consistent with that of observed x-ray
clusters. This limit on the systematic error in $H_0$ caused by cluster
shape assumes that each sample beta model cluster has fixed shape;
deviations from constant shape within the clusters may introduce additional
uncertainty or bias into this result.
\end{abstract}


%  Main Body

\section{Introduction}

There has been a substantial effort to detect the Sunyaev-Zel'dovich (SZ) 
effect from galaxy clusters (\markcite{Suny}Sunyaev \& Zel'dovich 1972) 
and to analyze its distortion of the cosmic microwave background radiation 
(CMB) in conjunction with cluster x-ray properties to derive the cluster 
cosmological angular-diameter distance and thus estimates of the cosmological 
parameters $H_0$ and $q_0$ (\markcite{Gunn78}Gunn 1978, 
\markcite{SilW78}Silk \& White 1978, \markcite{CavD79}Cavaliere, Danese, 
\& DeZotti 1979, \markcite{Birk79}Birkinshaw 1979; see also 
\markcite{Birk98}Birkinshaw 1998, 
and references therein). This method provides a distance determination for
the cluster that is independent of the ``cosmic distance ladder'' of 
Cepheid variables or supernov\ae, and is potentially effective for 
clusters at high redshift ($z \simeq 1$). Centimeter-wavelength interferometry
optimized for imaging the SZ effect from galaxy clusters 
has been recently developed (\markcite{Carl96}Carlstrom, Joy, and Grego 1996;
\markcite{Grai96}Grainge \etal 1996;). This allows 
high-resolution x-ray and radio images of clusters to be analyzed 
simultaneously. The results of fits of both the x-ray and radio images 
to simple cluster-plasma models will yield improved estimates of 
$H_0$, and  systematic errors in the measured value of $H_0$ are likely to
be a significant limit to its accuracy.

Sources of systematic errors in the SZ-determined $H_0$ and $q_0$
can originate from the assumptions made in modeling the cluster plasma: ignorance 
of the cluster 
plasma's true  three-dimensional distribution and inadequate 
treatment of the physical state of the cluster plasma. Radio and x-ray 
images only provide the projected x-ray surface brightness and CMB decrement. For 
the analysis to proceed some assumption must be made about the cluster size 
along the line of sight; \eg, one assumes that cluster has spherical or 
ellipsoidal symmetry. The modeling of physical state of cluster plasma for 
SZ analysis has generally assumed that the plasma was of a single phase and 
temperature, using the somewhat {\sl ad hoc} ``beta'' model for electron 
density, $n_e(\vecr) \propto 
(1 + r^2/r_c^2)^{-3\beta/2}$, where $r_c$ is cluster's ``core radius'', 
within which the density is relatively flat. The beta model can be argued as a 
possible distribution for the plasma in a dynamically relaxed isothermal cluster in 
hydrostatic equilibrium (\eg, \markcite{CavF78}Cavaliere \& Frusco-Femiano 1978; 
\markcite{Sara86}Sarazin 1986), but its usefulness is more empirical; many 
x-ray images of clusters fit a beta model reasonably well (\markcite{Mohr99}Mohr \etal 
1999; see \S4). Also, studies of the distribution of SZ systematic errors caused 
by cluster shape and orientation (and other effects) based on the results of a large ensemble of 
numerically simulated clusters have yet to be completed; current results are for
a small set of simulated clusters (see \S4). Thus, three-dimensional ``toy model'' estimates 
for the effects of cluster shape are a useful 
first step in estimating these errors, and can help identify the physical sources of 
bias and scatter in $H_0$ estimates from simulated clusters.

In this paper I study the systematic errors in the value
of $H_0$, measured by SZ and x-ray observations, caused by effects of cluster shape. 
This study consists of two parts. First, I create theoretical galaxy cluster samples,
where each cluster's plasma distribution follows a triaxial isothermal beta-model (\S2),
possessing three independent core radii. I use the triaxial beta model because it is a 
simple three-dimensional generalization of the spherical or ellipsoidal beta models 
(commonly used in SZ analysis) that demonstrates the effects of shape and orientation on
the uncertainties in $H_0$ determined by SZ observations. The triaxial beta model
also produces simple analytical functions for the CMB decrement and x-ray surface
brightness so results for large samples of clusters can be easily calculated.

I create numerical distributions clusters by uniformly and 
independently sampling the plasma core radii, constraining them by a minimum ratio between
any two core radii of a sample cluster. These samples are uniform in the plane of allowed
cluster oblateness and ellipticity. The clusters are placed in the sky
with a random orientation to our line of sight. I identify a cluster sample with an 
optimum asphericity that has a distribution of apparent cluster ellipticities that is consistent 
with that of observed x-ray clusters (\markcite{Mohr95}Mohr \etal 1995; see \S3). 

Second, I analyze the clusters of the theoretical sample to determine their distance as if 
they were either spherical or an ellipsoid of rotation, as in done in 
observational analysis (\eg \markcite{HuBi98}Hughes \& Birkinshaw 1998). An important 
unknown quantity is an ellipsoidal cluster's inclination angle 
$i$; the estimated value for $H_0$ will vary greatly with $i$. However, since our theoretical 
clusters are actually three-dimensional, specifying a single inclination angle is artifical. 
Therefore, I  analyze each cluster very simply as if its inclination angle 
$i = 90^\circ$, \ie, that the core radii for the clusters are not altered by 
projection effects, and then study the distribution of the estimates for $H_0$
for a large number of sample clusters. The apparent shape of a sample cluster's 
x-ray surface brightness will be elliptical, with a large
and small angular core radius, $\theta^+ \geq \theta^-$. I 
calculate two different estimates of $H_0$ 
which are proportional to either $\theta^+$ or $\theta^-$, 
designated $\hat H_0^+$ 
and $\hat H_0^-$. 
I also calculate an estimate $\hat H_0^{\rm avg}$, by 
using the arithmetic average of $\theta^+$ and $\theta^-$.
 
I find that the sample means of the estimates $\hat H_0^+$ and 
$\hat H_0^{\rm avg}$ fall within $\simeq 5\%$ of $H_0$ for the optimal 
sample. The sample distribution of $\hat H_0^-$ shows greatest bias, 
with a mean for $\hat H_0^-$ that underestimates $H_0$ by $\simeq 14\%$ for my 
optimal cluster sample. 

As a predictor for SZ observations, I also calculate estimates for $H_0$ averaged 
for 1000 realizations of a sample of 25 clusters.
I find that the systematic errors caused by cluster shape
are limited: the $99.7\%$ confidence intervals for $\hat H_0^+$ and
$\hat H_0^{\rm avg}$ include the assumed value of $H_0$ for
my optimal cluster sample, and do not extend beyond $\simeq 14\%$ from $H_0$. The 
$99.7\%$ confidence interval for $\hat H_0^-$ does not include $H_0$,
indicating that it may not be a useful parameter for distance estimation.

The structure of this paper is as follows. In section \S2 I describe the triaxial beta model for
the cluster plasma, and describe the analytic expressions for their CMB decrement and 
x-ray surface brightness. I also describe
the construction of samples of theoretical clusters, distinguished by their degree of
triaxiality, and describe the manner in which I analyze the apparent clusters to
determine values for $H_0$. I present our results in \S3, followed by a summary
and discussion -- noting some of the limitations of this beta model based analysis -- 
in \S4.

\section{Method}

\subsection{Triaxial beta model clusters}

The distribution of cluster plasma is described by an isothermal ``beta''
model. The  electron density at a position within the cluster $\vecr = 
x_1{\hat {\rm {\bf {x}}}}_1 + x_2{\hat {\rm {\bf {x}}}}_2 +
x_3{\hat {\rm {\bf {x}}}}_3$, measured in the observer's coordinates, is given
by

\beq \label{ed1}
n_e(\vecr) = n_{e0} \bigl (1 + \vecr \cdot \vecM 
\cdot \vecr)^{-{3 \over 2} \beta},
\eeq

where the matrix $\vecM$ describes a cluster's shape and orientation of
its principal axes to the observer, and $\beta$ is an exponent with the 
nominal range $\case{1}{2} < \beta \leq 1$. The maps of x-ray surface brightness $S_X$
and cosmic-microwave background decrement $\delta T_r$ in sky angular coordinates 
$(\vartheta, \varphi)$ (measured from the cluster center $\vecr = 0$) are given by 
integrals of the x-ray emissivity and electron pressure over the line-of-sight 
(defined here as $x_1$) though the cluster;

\beq \label{xsb1}
S_X (\vartheta, \varphi) = {{1}\over{4\pi (1+z)^3}} \, \int \, 
\Lambda_X (T_e(\vecr)) \, n_e^2 (\vecr) \> dl
\eeq

and

\beq \label{dtt1}
\delta T_r (\vartheta, \varphi) \equiv {{\Delta T_r}\over{T_r}} (\vartheta, \varphi) = 
-2 \, {{k_B \sigma_{T}}\over{m_e c^2}} \, \int \, T_e(\vecr) \,  n_e(\vecr) \> dl.
\eeq

Here $T_e$ is the electron temperature, hereafter assumed  to be constant, 
$\Lambda_X (T_e)$ is the plasma emission function over a prescribed x-ray 
bandwidth at temperature $T_e$, and $z$ is the cluster redshift.

For a triaxial isothermal beta model
plasma described by equation (\ref{ed1}), then integrating equation (\ref{xsb1}) by
choosing $dl = dx_1$ gives $S_X$ to be

\beq \label{xsb2}
S_X (\vartheta, \varphi)  = {{B({{1}\over{2}}, 3\beta - {{1}\over{2}})}\over{4\pi}}
 \; {{n_{0e}^2\Lambda_X(T_e)}\over{(1+z)^3}} \; { {L_{eff}} \over
{\chi(\vartheta, \varphi)^{ 3\beta - {{1}\over{2}} }} },
\eeq

where $\chi (\vartheta, \varphi)$ is a quadratic function of the sky angular 
coordinates $\vartheta$ and $\varphi$, describing elliptical isophotes. Along the 
line of sight of the center $\chi (0,0) = 1$. The quantity $B(q,r)$ is the beta function.
The quantity $L_{eff}$ is an effective column length for the plasma along the
line of sight through the cluster:

\begin{eqnarray} \label{leff}
L_{eff}^{^{\rm triaxial}} & = & {{1}\over{\sqrt{M_{11}}}} \nonumber \\
& = & \biggl 
\{ {{\cos^2 \alpha_1}\over{r_{c1}^2}} + 
\sin^2 \alpha_2 \bigl ( {{\cos^2 \alpha_3}\over{r_{c2}^2}} +
    {{\sin^2 \alpha_3}\over{r_{c3}^2}} \bigr ) \biggr \}^{-1/2}.
\end{eqnarray}

The quantities $(r_{c1}, r_{c2}, r_{c3})$ are the cluster core radii, and $(\alpha_1, 
\alpha_2, \alpha_3)$
are the rotation angles of the cluster principal axes
relative to the observer. The coefficients in the quadratic function $\chi (\vartheta, 
\varphi)$ are also functions of the cluster core radii and its orientation.

By integrating equation (\ref{dtt1}) in exactly the same manner, $\delta T_r$ can be shown to be

\beq \label{dtt2}
\delta T_r (\vartheta, \varphi) = - B({{1}\over{2}}, {{3}\over{2}}\beta - {{1}\over{2}}) \;
 n_{e0} \sigma_T \, {{k_B T_e}\over{m_e c^2}} \; {{L_{eff}} \over
{ \chi(\vartheta, \varphi)^{{{3}\over{2}}\beta  - {{1}\over{2}} }} }.
\eeq

The value of $L_{eff}$ determined from observations by relating $S_X$ and $\delta T_r$ 
observed for the
cluster. For example, $L_{eff}$ can be determined by the values of $\delta T_r$ and $S_X$ 
measured at the cluster's center:

\beq \label{leff2}
L_{eff} = {{B({{1}\over{2}}, 3\beta - {{1}\over{2}})}\over{B^2({{1}\over{2}},{{3}\over{2}}\beta 
- {{1}\over{2}})}}
{{1}\over{(1+z)^3}} {{\Lambda_X(T_e)}\over{4 \pi \sigma_T^2}} \biggl ( 
{{m_ec^2}\over{k_B T_e}} \biggr )^2
{{\delta T_r^2 (0,0)}\over{S_X (0,0)}}.
\eeq

The cluster's cosmological angular diameter distance $D_\theta (z; H_0, q_0)$ 
is then inferred by equating the measured $L_{eff}$ to that derived from a model
for the cluster. Recent efforts to fit the cluster 
$S_X$ (\markcite{HuBi98}Hughes and Birkinshaw 1998) 
assume that the cluster is either an oblate or prolate ellipsoid in shape as 
well as isothermal; this assumption about shape is reasonable when no information 
can be known about the structure of the cluster along the line of sight. However, the 
dependence of $L_{eff}$ on the cluster's apparent major and minor axes for an 
ellipsoidal beta model is different than that for a triaxial cluster.

\beq \label{leff3}
L_{eff}^{^{\rm ellipsoidal}} = D_\theta (z; H_0, q_0) \biggl 
\{ {{\cos^2 i}\over{\theta_{c2}^2}} +
          { {\sin^2 i}\over{\theta_{c1}^2} } \biggr \}^{-1/2},
\eeq

where $\theta_{c1}$ and $\theta_{c2}$ are an ellipsoidal clusters' angular axes; 
$\theta_{c1} > \theta_{c2}$ describes an oblate ellipsoid and $\theta_{c1} < \theta_{c2}$ describes a 
prolate ellipsoid. The quantity $i$ is the inclination angle of the symmetry axis.

If the cluster is triaxial, the observed $L_{eff}$ will be equal 
to that of equation (\ref{leff}). However, analyzing the cluster as an 
apparent ellipsoid with equation (\ref{leff3}) will produce a systematic error 
in the inferred value of $D_\theta$, and thus in cosmological parameters $H_0$ and $q_0$. 

\subsection{The theoretical cluster samples}

%I numerically generate two samples of 
%1000 clusters, one with $0.7 \, r_{c1} \leq r_{c2,3} \leq r_{c1}$, designated sample
%A; another with $0.5 \, r_{c1} \leq r_{c2,3} \leq r_{c1}$, is designated sample B.

I generate triaxial beta-model clusters choosing a set of core radii $(r_{c1}, r_{c2},
r_{c3})$ from a random uniform distribution for the ratio of two of the core radii, 
$r_{c2}$ and $r_{c3}$, with respect to $r_{c1}$. Both $r_{c2}$ and $r_{c3}$ are assumed to
be a random fraction of the length of $r_{c1}$, but bounded below by a minimum value. This 
minimum value is not known a priori but can be chosen to optimize the observed ellipticity of
x-ray clusters (see below). Clusters are only distinguished by their core radii; I do not 
create a distribution for the clusters' $\beta$-values nor any other quantity except core radii.
Spherical beta model fits to real clusters appear to exhibit correlation between 
core radius and the value of $\beta$ (\markcite{NeuA99}Neumann and 
Arnaud 1999). However, the 
observed correlation for beta model clusters is not convolved by
cluster shape projection. This is shown in the expressions for the
x-ray surface brightness $S_X$, equation (\ref{xsb2}), and SZ CMB
decrement, $\delta T_r$, equation (\ref{dtt2}). The profile exponents
for both of these quantities are not functions of the individual core radii
nor of rotation angles; nor are the major and minor axes of the
observed elliptical cluster ($\theta^+$ and $\theta^-$, see \S2.3) functions
of $\beta$; they are only functions of the sample cluster's core
radii and the rotation angles. Also, I am not considering a distribution of the
magnitude of the cluster core radii, but the distribution of the cluster triaxiality
(see \S2.3).

I rule out bias in the cluster samples toward a net oblateness or prolateness by checking
for uniform sampling in the ellipticity-prolateness $(E,P)$ plane, given by Thomas 
\etal (\markcite{Thom98}1998) as

\beq \label{edef}
E \equiv {{1}\over{2}} {{r_{c2}^2 (r_{c1}^2 - r_{c3}^2)}\over{r_{c2}^2 r_{c3}^2 + 
r_{c1}^2 r_{c3}^2 + r_{c1}^2 r_{c2}^2}},
\eeq

and 

\beq \label{pdef}
P \equiv {{1}\over{2}} {{r_{c2}^2 r_{c3}^2 - 2 r_{c1}^2 r_{c3}^2 + r_{c1}^2 r_{c2}^2}
         \over{r_{c2}^2 r_{c3}^2 + r_{c1}^2 r_{c3}^2 + r_{c1}^2 r_{c2}^2}}.
\eeq

Strictly prolate and oblate clusters fall onto the lines $P = -E$ and $P = E$ respectively,
with length determined by the lower bound of the ratio between core radii.
My optimal sample, described below, uniformly covers the allowed region in the 
$(E,P)$ plane, which is a triangle 
proscribed by the prolate and oblate lines and the line connecting their endpoints.

How well does the triaxial beta model reproduce the observed shapes of x-ray clusters that 
could be used for SZ analysis? A study of 65 {\it Einstein} x-ray clusters by Mohr \etal 
(\markcite{Mohr95}1995)
found an emission-weighted mean ellipticity of $0.20 \, \pm 0.12$, while 
McMillan, Kowalski, and Ulmer (\markcite{Mmku89}1989) found a mean ellipticity of $0.24 
\, \pm 0.14$ for 49 {\it Einstein} Abell clusters. In both of these studies 
clusters were included with substantial flattening caused by recent merging, or with 
cooling inflows which can make the cluster appear more spherical. A more appropriate sample for comparison 
would be one which excludes these effects. If I eliminate clusters that are apparent mergers
from the Mohr \etal sample (8 out of 12 clusters with ellipticities of $0.3$ or greater 
with apparent subclustering) and clusters in which cooling inflows may exist 
(as measured by central cooling times of 10 Gyr or less;
an additional 17 clusters), then the mean ellipticity of the remaining subset is $0.18 \, \pm 0.09$.
I find that a triaxial beta model cluster sample where the minimum ratio between any two
core radii to be $\simeq 0.65$ produces a distribution of apparent ellipticities that is
consistent with this subset of the Mohr \etal sample (figure 1). A Kolmogorov-Smirnov test 
between these samples indicates the ellipticity distributions are statistically indistinguishable,
with a maximum difference between the two cumulative distributions of $d = 0.12$, and a probability
that the two samples are drawn from the same distribution of $\simeq 69\%$.

\subsection{The analysis}

The systematic error in analyzing the clusters arises from assuming that
an apparent cluster is either a prolate or an oblate ellipsoid, when it is in fact
triaxial. The apparent elliptical image of the cluster will have a major and minor axis,
$\theta^+$ and $\theta^-$; these are the ``angular core radii'' for the x-ray and SZ images.
Using equation (\ref{leff}) and the function $\chi(\vartheta, \varphi)$,
$\theta^+$, $\theta^-$, and $L_{eff}$
are determined for a given triaxial cluster. The observational analysis proceeds as
if the observed $\theta^+$ and $\theta^-$ are that of an ellipsoidal cluster, 
inclined to the line of sight by an unknown angle $i$. The apparent 
cluster distance $\hat D_\theta$ for an ellipsoidal 
cluster is related to $\theta^+$, $\theta^-$, 
$L_{eff}$, and $i$ by deprojecting the cluster axes and using equation (\ref{leff3}):

\beq \label{mxxobs}
L_{eff}  = \hat D_\theta (z; \hat H_0, \hat q_0) \cases{ \sqrt{ {{\theta^{-^2} 
(\theta^{+^2} - 
\theta^{-^2} \cos^2 i)} \over {\theta^{+^2} \sin^2 i}} }, &{\rm (prolate)};\cr
              \sqrt{ {{\theta^{+^2} (\theta^{-^2} - \theta^{+^2} \cos^2 i)} \over 
{\theta^{-^2} \sin^2 i}} }, &{\rm (oblate).}\cr}
\eeq

Equating $L_{eff}$ of equation (\ref{mxxobs}) with that for the triaxial cluster, then
in general $\hat D_\theta$ will not be equal to the actual distance $D_\theta$. This leads to 
erroneous values for the apparent cosmological parameters $H_0$ and $q_0$.

Since the sample clusters are intrinsically triaxial, using estimators (\ref{mxxobs})
for $H_0$ based on an ellipsoidal cluster model with a single inclination angle $i$ is 
artificial; there is no single angle that characterizes the orientation of a cluster to
the line of sight, unless a sample cluster's core radii had been accidentally chosen 
to be roughly prolate or oblate. Therefore, I collapse the dependency on $i$ and
assume $i \equiv 90^\circ$, and use only a cluster's observed $\theta^+$ and $\theta^-$.
I compose three estimates for $H_0$ for each cluster: $\hat H_0^+ \propto 
\theta^+$, $\hat H_0^- \propto \theta^-$, and
$\hat H_0^{\rm avg} \propto \case{1}{2} (\theta^+ + \theta^-)$.
These estimates are ordered $\hat H_0^+ \geq \hat H_0^{\rm avg} 
\geq \hat H_0^-$. The estimate $\hat H_0^+$  
is equivalent to assuming the observed cluster an oblate ellipsoid, while the 
estimate $\hat H_0^-$ is for the cluster as a prolate ellipsoid, both viewed
as if the axis of rotation were in the plane of the sky.
I study the distribution of these estimators for $H_0$ for a sample of 
triaxial clusters drawn from the distribution described in \S$2.2$. 
The cosmological parameter $q_0$ is fixed to be zero, 
so that $\hat D_\theta = c \hat H_0^{-1} z (1 + \case{1}{2} z) / (1 + z)^2$. 
As mentioned in \S2.2, I am sampling clusters only by a distribtution in 
their triaxiality, and I do not
use the magnitude of the core radii. Thus the estimators $\hat H_0$ are 
determined with respect to an assumed value of $H_0$.

\section{Results}

Figure 1 shows the distributions of the values of $\hat H_0^+$ and
$\hat H_0^-$ inferred from the cluster apparent angular axes, a scatterplot 
of these estimates (one point per cluster), and the distribution of apparent 
cluster ellipticities, for our optimal theoretical beta model sample (with a 
minimum ratio of core radii of $0.65$). Figure 2 shows the distribution of
of $\hat H_0^{\rm avg}$ for this cluster sample.

The distributions of $\hat H_0^+$ has sample mean 
that is within $\simeq 8\%$ of the assumed value of $H_0$. The distribution of 
$\hat H_0^-$ is more biased (\ie, lower) in sample mean value, 
which is expected since for every sample
cluster the value of  $\hat H_0^-$ is constrained to be lower in value than
$\hat H_0^+$. 

I also studied the distribution of the $H_0$ estimates for clusters samples with greater
asphericity. For a cluster sample with a minimum ratio of core radii of $0.5$, the estimate 
distributions broaden significantly, and the sample deviations nearly double.
The distribution of $\hat H_0^{\rm avg}$ also exhibit broadening with 
greater asphericity, but still has a mean value within $\simeq 5\%$ of $H_0$. 
All of the means are relatively insensitive to the samples' degree of asphericity. 

What then is the expected systematic error in the measured $H_0$ caused by cluster shape for a 
practical-sized sample of clusters? The $99.7\%$ confidence intervals for the 
mean values $\hat H_0^+$, $\hat H_0^-$, and $\hat H_0^{\rm avg}$, 
based on 1000 realizations of a 25-cluster sample are summarized in Table 1. 
The confidence intervals for $\hat H_0^+$ and
$\hat H_0^{\rm avg}$ include the assumed value of $H_0$ for
the optimal cluster sample, and even for samples with even greater asphericity. 
For the optimal cluster sample the confidence intervals for $\hat H_0^+$ and
$\hat H_0^{\rm avg}$ do not extend beyond $\simeq 14\%$ from $H_0$. The 
$99.7\%$ confidence interval for $\hat H_0^-$ does not include the assumed
value of $H_0$.

\begin{table*}
\begin{center}
\begin{tabular}{cccc}
Minimum ratio of core radii &$\hat H_0^+/H_0$ &$\hat H_0^-/H_0$ &$\hat H_0^{\rm avg}/H_0$\\
\tableline
0.65 &0.94 - 1.14& 0.78 - 0.97& 0.87 - 1.05\\
\end{tabular}
\end{center}
\caption{$99.7\%$ confidence intervals for 25-cluster sample mean of $\hat H_0$. The mininum
ratio between core radii is $0.65$}
\end{table*}

\section{Summary and Discussion}

The high-resolution imaging of the SZ effect in galaxy clusters in
combination with cluster plasma x-ray diagnostics is a powerful
technique for measuring the cosmic distance scale.
This method is sensitive to the projection of the cluster's  
inverse-Compton scattering and x-ray emission, which 
depend on plasma density and temperature along the cluster
line-of-sight.

In this paper I have estimated the systematic errors in the SZ-determined
Hubble constant caused only by the projection effects of cluster shape. I use a
triaxial beta model to represent the clusters' gas because it is the simplest 
generalization of the ubiquitous spherical and ellipsoidal beta models that 
can demonstrate the effects of cluster shape and orientation on 
measurements of $H_0$. 
The triaxial beta model has analytic expressions for the 
clusters' CMB decrement and x-ray surface brightness, so that the 
statistics for measured $H_0$ for a very large number of 
clusters are easily computed. 

Ideal clusters for SZ analysis would possess no obvious 
substructure nor evidence of merging, and contain plasma at a 
single temperature without cooling inflows. As most observed clusters are
not such simple systems, I discuss the relevance of 
my beta model sample predictions for errors in $H_0$ caused
by cluster shape for a real SZ cluster survey. First, will
the presence of cooling gas alter a cluster's SZ properties 
substantially from a beta model description ?  A recent analysis 
of \ROSAT x-ray clusters indicate that a majority of them contain
cooling gas (\markcite{Pere98}Peres \etal 1998).
The clusters' x-ray emission is sensitive to the environment of their centers, where
gas densities are highest, while their SZ effect is relatively more
sensitive to the lower-density outer regions of the clusters. 
Cooling inflows are confined to the core of the clusters and 
are believed to be subsonic and isobaric, so that the relevant 
SZ cluster property, the column integral of the cluster pressure, will be 
largely unaffected with the presence of cooling gas in the core.
``Reprojection'' estimates for \Einstein x-ray clusters suggest that
the SZ effect is enhanced in the largest of cooling clusters,
with cooling rates of hundreds of solar masses per year 
(\markcite{Whit97}White \etal 1997). However, most of these clusters have 
high pressures in their outer regions, so the SZ enhancement is likely not to be
caused by the alteration of the SZ effect within the cooling core, 
but by the presence of higher gas pressure over 
the extended outer (beta model) region. Strong cooling inflow clusters have sharply peaked
x-ray profiles so that their x-ray determined core radii and central plasma
densities can be skewed. 
As mentioned in \S2.2, in comparing my beta model sample
to the observed cluster ellipticity distribution, I eliminate the stronger cooling
inflow clusters in attempting to avoid this bias in observed shape.
For many of these systems these issues are academic, as they
are radio loud and obscure a CMB decrement (\markcite{Burn90}Burns 1990).
However, some of the clusters that have a detected SZ effect contain large cooling
inflows (\markcite{Hugh97}Hughes 1997), and observers have used the beta model 
properties of the outer portion of the clusters to extract the relevant x-ray properties
for analysis in conjunction with the SZ effect (\eg, \markcite{Myer97}Myers \etal 1997).  
It is also interesting to note that for many clusters -- approximately half of \ROSAT clusters 
of an x-ray flux-limited sample previously selected by 
Edge \etal (\markcite{Edge90}1990), including 
some that contain cooling inflows -- the simple beta model can produce a good fit to 
their x-ray profiles (\markcite{Mohr99}Mohr \etal 1999).

An important caveat to the error estimates in $H_0$ provided by these  
beta models is that they cannot account for the cluster gas distibutution 
changing shape from the core to the outer region.
Observations of a few high signal to noise \ROSAT clusters show ellipticity 
gradients, exhibiting a rougly linear decline in x-ray ellipticity from 
$e \simeq 0.3 - 0.1$ 
from the clusters' center, over a distance of several core radii
(\markcite{Buot96}Buote and Canizares 1996). This behavior may have a substantial 
effect on the SZ properties of a cluster, altering both the shapes and magnitudes
of the apparent x-ray and (to an even larger degree) the SZ images. 
%While $99\%$ of the central x-ray 
%surface brightness for a $\beta = 2/3$ model cluster comes from gas 
%within $\simeq 8$ core radii, gas outside of this region contributes $\sim 25\%$ 
%of the central CMB decrement.
While results of cluster ellipticities
from a larger sample are required to adequately constrain this effect 
for a statistical study, here I illustrate a possible bias in $\hat H_0$
that could arise from changing cluster shape using 
a simple model of a cluster with varying ellipticity.
I consider a beta model with the core radii in the elements of 
$\vecM$ in equation (\ref{ed1}) as functions of coordinates. 
An {\sl ad hoc\/} example is 
\beq \label{rcv}
r^2_{c1} \rightarrow r^2_{c1} + 
\biggl [ {{r}\over{r_{c1} + r }} 
\biggr ]^\alpha ( R^2 - r^2_{c1} ),
\eeq
used in the beta model of equation (\ref{ed1}), with
where $r$ is the distance of the coordinate point from the cluster center.
This describes a triaxial cluster (using similar expressions for 
$r_{c2}$ and $r_{c3}$) with core radii of $(r_{c1}, r_{c2}, r_{c3})$ 
within the cluster core, becoming spherical with core radius $R > r_{c1}$ outside 
the cluster's core. I refer to this as a cluster with a ``modified'' core radius.
Choosing $\alpha \simeq 4$ and $r_{c1}/R \simeq 0.6$
for a cluster with one core radius modified by equation (\ref{rcv}) and 
$\beta = \case{2}{3}$ in equation (\ref{ed1}), produces a decreasing x-ray 
ellipticity from $e (R) \simeq 0.3$ to $e (5R) \simeq 0.1$, when the modified
core radius lies in the plane of the sky; there is similar behavior in ellipticity for two 
modified cluster core radii with one along the line of sight. The presence of gas with
a more spherical distribution moderates the effect of cluster orientation on $L_{eff}$
determined from equation (\ref{leff2}), with $L_{eff}$ assuming intermediate values 
within the range given by the outer core radius $R$ and the inner core radii
$(r_{c1}, r_{c2}, r_{c3})$. For example, a cluster with one modified core radius, taken
along the line of sight and using the values for $r_{c1}/R \simeq 0.6$ 
and $\alpha \simeq 4$, produces an estimate $\hat H_0$ that is biased (high) 
by $\simeq 20\%$.
This by itself is a substantial effect on what otherwise appears as
a spherical cluster, however in conjunction with orientation, it is a 
significantly lower bias that would have been produced by the unmodified oblate
cluster observed along its minor axis, $\simeq 67\%$. I have calculated the
x-ray and CMB decrement images for a small set of
prolate or oblate clusters, with one or two modified core radii
(using the parameters from above), observed
along the axes and along the line $x = y = z$. For these clusters the estimates 
$\hat H_0^+$ and $\hat H_0^-$ are lower (by $\approx 10\%$), and their 
difference $\hat H_0^+ - \hat H_0^-$ 
substantially smaller than their counterparts for clusters with unmodified core radii.
It is uncertain whether any significant bias to estimates for $H_0$ would be introduced
in statistically analyzing a large set of such type of clusters. 
These estimates are based on using the central values for x-ray brightness, CMB decrement, 
and determining the
cluster angular size using the apparent x-ray core radii defined simply by the ellipse 
of brightness that is lower than the peak by a factor $2^{3\beta - 1/2}$ with
$\beta = \case{2}{3}$. 
More quantitative results will depend on the model details of the distribution of gas in 
transition from core to outer region of the cluster, and the manner by which models
are fit to
the data to determine parameters (\eg simultaneous x-ray and CMB image fitting),
well beyond the scope of this paper. Qualitatively, however, the
presence of a changing cluster shape can alter the estimates for
$H_0$ by softening the effects of orientation. A notable effect for this type of 
cluster is that the beta model congruence of
the x-ray and CMB images (equations (\ref{xsb2}) and (\ref{dtt2})) is broken,
so that comparison of the maps may determine the importance of shape
changes outside of the cluster core.
 
Clusters with recent merging activity cannot be adequately represented by 
simple beta models. However, the use of such clusters in a SZ survey is likely to be
fraut with complications. In principle, numerical simulations of 
cluster formation would yield a more ``realistic'' sample of 
clusters for SZ analysis than my beta model sample, accounting 
for the effects of cooling and merging as well as for shape projection.
However, simulations of cluster formation can not yet physically 
reproduce the observed large gas cores that are observed in clusters 
(\markcite{MetE97}Metzler and Evrard 1997; \markcite{AnnN96}Anninos and Norman 1996).
\markcite{Inag95}Inagaki \etal (1995) 
used two simulated clusters to study SZ measurement systematics
caused by plasma temperature gradients, plasma clumpiness, 
cluster peculiar velocity, the finite extent of cluster plasma, 
and cluster shape. They determined that effects of asphericity
would be limited to an uncertainty of $\simeq 10\%$ in $H_0$
by observing several clusters, as I have also found.
They did not conduct a survey of possible cluster shapes; the 
statistics for estimates of $H_0$ were generated by
the viewing of the two clusters at many orientations. 
I have constrained the limits of my beta model
shapes by checking for its consistency with the
observed apparent ellipicities of x-ray clusters.
\markcite{Roet97}Roettiger \etal (1997) focused on
the systematic errors in an SZ-determined $H_0$ observed
in seven simulated cluster mergers with strong temperature gradients and 
asphericity. They
found that these effects could lead to $H_0$ underestimated 
by as much as $35\%$, and concluded that two approaches should
be used in SZ analysis: perform detailed simulations of 
individual clusters where it was indicated that merging 
was strongly affecting the SZ properties, otherwise employ
a statistical sample of clusters that show no evidence of
recent merging or dynamical evolution. In this paper I have 
addressed the systematic errors caused by cluster shape and 
orientation that would be present in using this latter 
approach with a modeled optimal SZ cluster sample. What are
needed now are the statistical results for SZ $H_0$ estimates 
from a large sample of numerically simulated clusters.

I have created numerical samples of triaxial beta model clusters by
specifying the minimum ratio between any two core radii. I have 
identified an optimal such sample, with the ratio of $\simeq 0.65$, 
that has a distribution of apparent cluster x-ray ellipticities that 
is consistent with that measured from observations of x-ray clusters.
I have analyzed the cluster samples for their SZ decrement and
x-ray surface brightness, assuming no effects of inclination angle.
The apparent cluster's large and small angular core radius, $\theta^+$ 
and $\theta^-$, yield three estimates of $H_0$ that are proportional 
to $\theta^+$, $\theta^-$ and $\case{1}{2} (\theta^+ + \theta^-)$. 
These estimates are equivalent to assuming that the cluster is either oblate 
($\propto \theta^+$) or prolate ($\propto \theta^-$), 
with its symmetry axis in the plane of the sky, or spherical 
$\propto \case{1}{2} (\theta^+ + \theta^-)$.
I have found that the estimates $\hat H_0^+$ and $\hat H_0^{\rm avg}$, 
have means that fall within $\simeq 5\%$ of the assumed value of $H_0$ 
for the optimal theoretical cluster sample, while the mean of the estimate 
$\hat H_0^-$ underestimates $H_0$ by $\simeq 14\%$. 
The size of these errors caused by cluster shape is similar to that found 
in a more approximate fashion by Hughes and Birkinshaw (\markcite{HuBi98}1998),
and discussed recently by Cooray (\markcite{Coor98}1998). 
Other estimates of $H_0$ can be devised, for example, a (weighted) geometric mean
$\tilde H_0 \equiv (\hat H_0^+)^\alpha (\hat H_0^-)^{1-\alpha}; 
\, 0 < \alpha < 1$ (\markcite{VanV97}Van Speybroeck and Vikhlinin 1997). 
These may produce better estimates for $H_0$ than the three simple estimates 
that I have studied, but the best choice of $H_0$ estimator may depend on 
the intrinsic shape distribution of clusters.

I have also determined the confidence intervals for the estimates of $H_0$
that would be derived from the SZ and x-ray analysis of a 25-cluster sample. 
Our optimal theoretical cluster sample has $99.7\%$ confidence intervals 
for $\hat H_0^+$ and $\hat H_0^{\rm avg}$ that are within 
$\simeq 14\%$ of $H_0$, and enclose $H_0$. The confidence intervals 
for the estimate $\hat H_0^-$ show more deviation and do not enclose $H_0$, 
indicating that it may not a useful estimator. 

\acknowledgments

I thank J. Mohr, A. Evrard, and B. Mathiesen for very enlightening conversations
and suggestions. I thank M. Joy and S. Patel for critical readings of earlier
versions of this manuscript. I also thank NASA's Interagency Placement Program, 
the University of Michigan Department of Astronomy, and the University of Michigan 
Rackham Visiting Scholars Program.

%\begin{figure}
%\plotone{fig01.ps}
%\caption{Fig 1: The coverage of model cluster samples on the $(E,P)$ plane. Our optimal
%sample has a minimum ratio of any two core radii of $0.65$. The
%triangle that proscribe the points is the region allowed in the $(E,P)$ plane for the given 
%limit on the axial ratio.}
%\end{figure}

\begin{figure}
\plotone{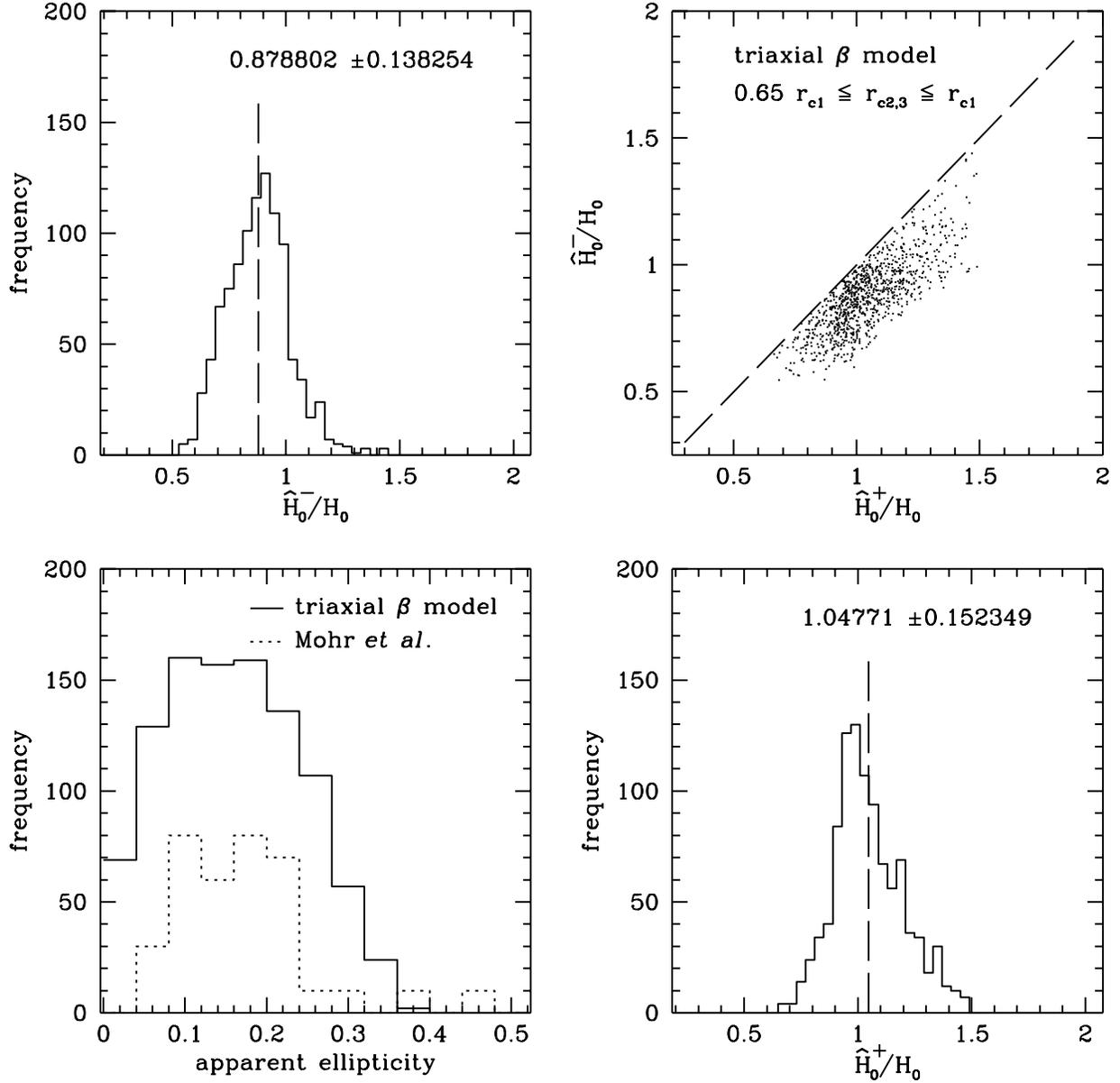}
\caption{The distributions of $\hat H_0$ inferred from the
cluster apparent core radii, compared to the assumed value of $H_0$, a scatterplot 
of $\hat H_0^+$ vs. $\hat H_0^-$, and the distribution of 
apparent cluster ellipticities, for the optimal sample of 1000 triaxial beta model clusters
This sample produces a distribution of apparent cluster
ellipticities that is consistent with the subset of
the Mohr \etal sample (see text).
Distribution mean values are indicated by dashed lines
(value of sample mean and deviation above line); the dashed line on the scatterplot 
indicates where cluster appears circular. The dotted-line histogram is that of the subset of the Mohr
\etal x-ray clusters described in the text (full scale for its frequency is $0-20$).}
\end{figure}

\begin{figure}
\plotone{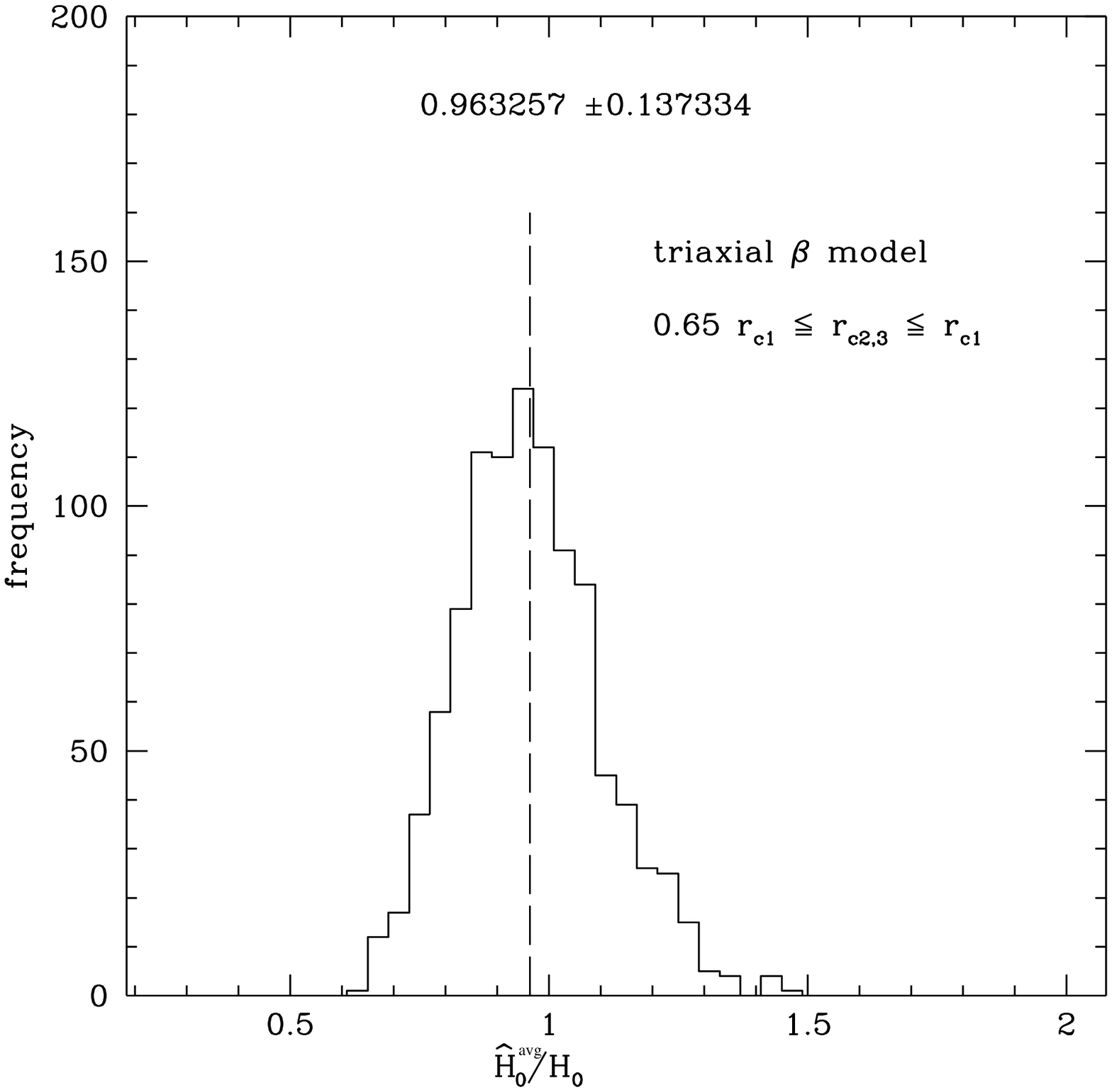}
\caption{The distributions of $\hat H_0^{\rm avg}$, inferred from the
clusters by averaging major and minor axes, compared to the assumed
value of $H_0$, for the optimal sample. The distribution mean values 
are indicated by dashed lines
(value of sample means and deviations above lines).}
\end{figure}


\begin{references}

\reference{AnnN96}Anninous, P. \& Norman, M.L., 1996, \apj, 459, 12
\reference{Birk98}Birkinshaw, M. 1998, {\rm Physics Reports}, in press
\reference{Birk79}Birkinshaw, M., 1979, \mnras, 187, 847
\reference{Buot96}Buote, D.A., \& Canizares, C.R., 1996, \apj, 457, 565
\reference{Burn90}Burns, J.O., 1990, \aj, 99, 14
\reference{Carl96}Calstrom, J.E., Joy, M. K. \& Grego, L. 1996, \apj, 456, L75
\reference{CavD79}Cavaliere, A., Danese, L. \& De Zotti, G., 1979, \aap, 75, 322
\reference{CavF78}Cavaliere, A. \& Fusco-Femiano, R., 1978, \aap, 70, 677
\reference{Coor98}Cooray, A.R., 1998, \aap, in press, astro-ph 9808186
\reference{Edge90}Edge, A.C., Stewart, G.C., Fabian, A.C., \& Arnaud, K.A., 1990,
\mnras, 239, 559
\reference{Grai96}Grainge, K., Jones, M., Pooley, G. Saunders, R., Baker, J., Haynes, T. 
\& Edge, A., 1996, \mnras, 278, L17
\reference{Gunn78}Gunn, J.E., 1978, {\rm In Observational Cosmology 1; eds Maeder, A., 
Martinet, L. \& Tammann, G.; (Sauverny: Geneva Observatory)}
\reference{Hugh97}Hughes, J.P., 1997, in Proceedings IAU 188 "The Hot Universe",
astro-ph/9711135
\reference{HuBi98}Hughes, J.P. \& Birkinshaw, M. 1998, \apj, in press, astro-ph/9801183
\reference{Inag95}Inagaki, Y., Suginohara, T. \& Suto, Y. 1995, \pasj, 47, 411
\reference{Mmku89}McMillan, S.L.W., Kowalski, M.P. \& Ulmer, M.P., \apjs, 70, 723
\reference{MetE97}Metzler, C. A. \& Evrard, A. E., 1997, astro-ph 9710324
\reference{Mohr99}Mohr, J.J., Mathiesen, B., and Evrard, A.E., 1999, \apj, in press, astro-ph/9901281
\reference{Mohr95}Mohr, J.J., Evrard, A.E., Fabricant, D.G. \& Geller, M.J., 1995,
\apj, 447, 8
\reference{Myer97}Myers, S.T., Baker, J.E., Readhead, A.C.S., \& Leitch, E.M., 1997,
\apj, 485, 1
\reference{NeuA99}Neumann, D.M, \& Arnaud, M., 1999, astro-ph/9901092
\reference{Pere98}Peres, C.B., Fabian, A.C., Edge, A.C., Allen, S.W., Johnstone, R.M., \&
White D.A., 1998, \mnras, 298, 416
\reference{Roet97}Roettiger, K., Stone, J.M. \& Mushotsky, R.F. 1997, \apj, 482, 588
\reference{Sara86}Sarazin, C. 1986, {\rm Rev. Mod. Phys.}, 58, 1
\reference{SilW78}Silk, J.I. \& White, S.D.M., 1978, \apjl, 226, L3
\reference{Suny72}Sunyaev, R.A. \& Zel'dovich, Y.A. 1972, {\rm Comm. Astrophys. Sp. Phys.}, 4, 173
\reference{Thom98}Thomas, P. A., Colberg, J.M., Couchman, H.P., Efstathiou, G.P.,
Frenck, C.S., Jenkins, A.R., Nelson, A.H., Hutchings, R.M., Peacock, J.A.,
Pearce, F.R. \& White, S.D.M. 1998, \mnras, 296, 1061
\reference{Whit97}White, D.A., Jones, C., \& Forman, W., 1997, \mnras, 292, 419
\reference{VanV97}Van Speybroeck, L. \& Vikhlinin, A., 1997, private communication

\end{references}
\end{document}